# The Effect of Problem Format on Students' Responses

Beth Thacker, Ganesh Chapagain, David Pattillo, Keith West, Physics Department, MS 41051, Texas Tech University, Lubbock TX 79409

## Abstract

As part of large-scale assessment project at Texas Tech University, we studied the effect of problem format on students' responses to quiz questions. The same problem was written in multiple formats and administered as a quiz in the large introductory physics sections in both the algebra-based and calculus-based classes. The formats included multiple-choice (MC) only, MC and free response (FR) and FR only. Variations in the FR wording were also explored. We examined the ability of students to both choose the correct answer and correctly explain their reasoning and show their calculations. We also analyzed the type of written responses the students used to support their answers. We found that a large percentage of the students who chose the correct answer could not support their answer with correct reasoning or calculations.

## I. INTRODUCTION

Multiple-choice (MC) assessment is a common method of assessment at large universities. It is often used as the dominant or exclusive form of assessment on exams and many times it is the case that students never have to show their calculations or explain the reasoning for their answers in any part of the course. The homework is often online and also does not require a demonstration of the process of determining an answer by a consistent, coherent argument or calculation. There are simultaneously instructors, even at large universities, who still require students to work at least some free-response (FR) (also known as constructed response (CR)) questions on exams and homework that require hand-grading by instructors and/or teaching assistants. While there has been research done on the benefits, drawbacks and comparisons of MC, FR and various hybrids and variations of the two assessment formats,[1-14] the research is by no means definitive. It is clear that tests can be designed so that there is a high correlation between MC and FR when tests are scored simply for the number of correct responses. (This means that if your goal is simply to distinguish A, B and C students, a well-designed MC exam can be as effective as FR at achieving your goal.) However, research on the equivalence of traits assessed, skills measured, level of complexity, ability to inform instructors on students strengths and weaknesses and ability of the assessment to inform instructional methods and curriculum development all need further and ongoing research. Students attitudes toward different assessment formats must also be researched, as that will affect students' motivations and study methods.

While some instructors are satisfied with the knowledge that the results of MC and FR testing can be highly correlated when exams are scored simply for the number of correct answers, others still favor FR testing, as it is claimed to be better at assessing thinking processes, higher level thinking skills, and informing instruction. There are many papers that demonstrate advantages of FR over MC.[8-14] These papers claim that FR is more reliable and valid, more discriminating and better at predicting future performance at school or work, among other things.

The fact that it often seems as if there are two camps of instructors, those who would argue for MC and those who are staunch supporters of FR, leads one to question the conclusiveness of the research. While it does depend on how you use and interpret the research and the goals of your assessment, there is still need for further research on assessment. Some of that need is arising due to the development of more student-centered curricula and non-traditional teaching methods designed based on research. Instructors teaching non-traditionally are often questioning the most appropriate form of assessment for the skills (and content) they are attempting to teach and are much more interested in how the assessment can inform instructors and instruction based on evidence of the students' thinking processes and cognitive abilities. Further research, particularly on the claims of the benefits of FR testing, is very much needed.

Indeed, based on the current research, there is a prevailing sentiment among many in both the MC and FR camps that FR is at least partially favored, but that since it is not cost and time effective, it is necessary to use MC and design it to give indistinguishable results from FR when scoring the number of correct responses. We believe that submission to this sentiment is hindering much needed research on FR testing that is particularly needed by those wishing to assess students taught in non-traditional formats.

As research on assessment has to take place in the context of a subject and an educational setting, we direct our research to the assessment of physics skills and concepts in introductory physics classes at the university level.

Examining the existing physics education research (PER), we find at least one study that indicates that it is possible to construct valid and reliable multiple-choice questions that correlate with FR testing when the number of correct responses is scored,[7] consistent with previous research results. However, there have also been studies that demonstrate evidence of significant differences in students' responses between multiple-choice and free response versions of the same problem,[15] that multiple choice questions can be false indicators of students' understanding[16] and that a significant number of students have alternative conceptions inconsistent with the distractors,[17] even on valid and reliable conceptual inventories.

This paper expands on this research. We report on a study of the effect of problem format on students' understanding in the context of a large-scale assessment project at a large university, looking not at exams as a whole (not at simply the final scoring), but at student responses to individual problems.

At Texas Tech University (TTU), we had been working, with the support of a National Institutes of Health (NIH) Challenge grant,[18] to assess the implementation of PER-informed methods and curricula in the laboratories and recitation sections of the introductory physics courses. We are at an institution where the majority of instruction is traditional lecture-based instruction and in large classes the majority of testing is in multiple-choice format and most of the homework is online. The project included the use of a number of different assessment instruments, including conceptual inventories, such as the Force Concept Inventory (FCI),[19] Scientific Attitude and Scientific Reasoning Inventories[20], TA Evaluation Inventories[21] and a set of locally written FR pre- and post-tests. It was with the introduction of the FR pre- and post-tests that we, for the first time, required students in the large classes to show their work and explain their reasoning on graded assessments.

As we analyzed the FR post-tests (graded quizzes), we observed that it was not unusual for a large percentage of students to be able to choose the correct answer, but not be able to correctly explain their reasoning or show their calculations. While it was problem dependent, as many 30% (or more) of the students would be able to choose the correct answer but not be able to provide a correct explanation. We observed this on a number of quizzes. We decided to use one of the quizzes each semester to study the effect of problem format on students' responses. We report on the results of two different quizzes administered in different problem formats in both the algebra-based and calculus-based introductory physics recitation sections. Both problems required either implicit or explicit ranking of the results of different scenarios. The formats included MC only, MC/FR, and FR only, with various wording changes and requirements written into the stem of the problem.

We report on the results of each of the problems, focusing on the comparison of MC, MC/FR, and FR formats as far as correct answers, but also on the type of incorrect answers provided by students in FR and MC/FR formats. In Section II, we describe the student populations, in section III, we discuss the question formats and the problem administration, in Section IV, we present the results and in Section V, we discuss and conclude.

## II. STUDENT POPULATIONS

The post-tests were administered as quizzes to two groups of students, the calculus-based and algebra-based introductory physics students, in Fall 2010, Spring 2011 and Fall 2011.

### A. Calculus-based Courses

The calculus-based class consists primarily of engineering and computer science majors. The number of students registered for the course is usually around 500, with 513 in Fall 2010, 527 in Spring 2011 and 471 in Fall 2011, split each semester among three lecture instructors. The instruction is primarily traditional lecture, with one one-hour recitation section and one two-hour lab. In these three semesters, the recitation and lab were taught in three consecutive hours, with one-hour recitation and two hours lab. The labs and recitations are common among the three instructors each semester. Students from each of the lecture instructors are mixed in the labs and recitations.

Most of the instructors taught in traditional lecture format. One instructor, who taught both in Fall 2010 and Spring 2011 was self-identified as using some interactive –engagement techniques in the lecture. There were no significantly different results among the three instructors each semester.

The recitations and labs were taught by teaching assistants (TAs). The TAs are trained in the lab and recitation preparations to use interactive-engagement techniques and student-centered instructional methods. However, the use of these techniques, as well as the extent of the use of more traditional teaching techniques, varies widely among TAs, based on formal[21] and informal observations.

### B. Algebra-based Courses

The algebra-based class consists mostly of pre-health science majors, including pre-medical, pre-dental, pre-physical therapy, etc. The number of students registered each semester is usually around 250-300, with 257 in Fall 2010, 276 in Spring 2011 and 227 in Fall 2011, split each semester among two lecture instructors. The instruction is primarily traditional lecture, with one one-hour recitation section and one two-hour lab. As in the calculus-based classes, in these three semesters, the recitation and lab were taught in three consecutive hours, with one-hour recitation and two hours lab. The labs and recitations are common among the three instructors. Students from each of the lecture instructors are mixed in the labs and recitations.

Most of the instructors taught in traditional lecture format. One instructor, who taught in Fall 2010 used interactive–engagement techniques. There were no significantly different results among the two instructors each semester.

As in the calculus-based classes, the recitations and labs were taught by TAs trained to use interactive-engagement techniques and student-centered instructional methods. Also, as in the calculus-based classes, the extent of the use of more traditional teaching techniques varies widely among TAs, based on formal[21] and informal observations.

## III. THE PROBLEM ADMINISTRATION AND PROBLEM FORMATS

### A. Problem administration and analysis

We report on two problems that required either implicit or explicit ranking, as written into the stem of the problem. The problems were written in MC, MC/FR and FR formats and in six (Problem A) and three (Problem B) different versions. The versions differed in implicit or explicit ranking, different wording in the stem of the problem and different requirements for the work or calculations to be shown. The problems are given in Appendices I and II, along with the research rubrics used to evaluate the students' answers. The first problem, in Appendix I, administered in Fall 2010 and Spring 2011, is a problem taken from the University of Illinois Physics Education Research (PER) website[22]. We will refer to it in the rest of this paper as Problem A.  The second problem, in Appendix II, administered in Fall 2011, is a problem similar to those in *Ranking Task Exercises in Physics: Student Edition*[23]. It will be referred to as Problem B in the rest of this paper. The content of the problems involves the concept of conservation of energy and the understanding and use of kinematics equations. The problems were administered as quizzes in the recitation sections after the content and skills had been covered in lecture, lab and recitation.

The problems were administered in the recitation sessions by the TAs at the beginning of the sessions. Each semester, three different versions of the quiz, on different colored sheets of paper, were randomly handed out to the students in each of the recitation sections. The students had fifteen minutes to complete the quiz. Students who were not present at the recitation or came late to the recitation session were not allowed to take the quiz.  The TAs collected the quizzes, turned them in to the researchers for scanning, and then graded the quizzes, after they had been scanned. The quizzes were graded by the TAs by a rubric determined by the lab/recitation coordinator each semester, which was not necessarily the rubric used for research analysis given in the Appendices of this paper.

The researchers analyzed the quizzes by the rubrics given in the Appendices.
The answers were classified into four categories: i) completely correct (CC), ii) correct choice, partially correct explanation (CP), iii) correct choice, incorrect explanation (CI) and iv) Incorrect (I).  The researchers also analyzed the CP, CI and I answers for components of critical thinking, such as the type and consistency of the arguments used and common incorrect conceptions.

## B. Problem Format Survey

Since we are aware of the strong views of many faculty regarding assessment format, including a prevalent view that FR, even if it is preferred, may not be the assessment of choice because it is not cost and time effective, we wondered about the students' opinions on testing format. Certainly a student's opinion on testing format would frame their understanding of course and instructor expectations and their study methods. While there exists "common" or "folk" knowledge about students' assessment preferences, there is surprisingly little research on the subject.[24-27] It is generally believed that students prefer MC, and there is some research supporting this.[24-25] We decided to ask the students at our university about their assessment preferences and thoughts on different assessment formats, particularly as far as how well they felt it represents their knowledge of the subject.

We administered a survey on problem format to 91 students in the calculus-based class in Spring 2012. The survey is given in Appendix III. We asked the students to identify the quiz or exam question format they preferred to be graded on, the format they thought would be easiest and the format they thought would give their instructor the most information on their understanding of the physics content. They were also asked to support their reasoning with explanations. We report the results in Section IV.C.ii.

## IV. RESULTS

## A. Results of written responses Problem A

As stated previously, we decided to focus on two individual problems administered in different formats as representative of the effects we were seeing when FR format quizzes were introduced into courses assessed predominantly by MC assessment instruments. We address both the numerical analysis (the percentage of students answering in each of four different categories, CC, CP, CI and I) and common CP, CI and I solution types.

## 1. Numerical analysis

The results for the calculus-based sections for Problem A are shown in Table 1 and Figure 1. The data is shown for each of the six formats in which it was administered, three in Fall 2010 and three in Spring 2011.

In Figure 1. the first graph in the upper left corner shows how students responded to the problem when it was administered in MC format. Sixty-nine and fifty-five percent of the students were able to choose the correct answer in part (a) and part (b), respectively. In this format, the students are not required to explain their answer or show their calculations. In any of the problems in FR or MC/FR format, the students are required to show their work and explain their reasoning. In these problems, the percentage of students answering completely correctly (CC) drops dramatically. The most striking feature, however, in the FR formats is the number of students who can choose the correct answer, but not correctly explain their reasoning or show their calculations (CI). This ranges from about 25 – 60% in part (a) and is around 60 – 70% in part (b).

Table 2 and Figure 2 show the same results for the algebra-based course. The results are very similar, with a significant drop in completely correct answers and a high percentage of students who can choose the correct answer and not correctly explain it (CI) when the problem is presented in FR format.  The percentage of CI answers ranges from about 15 – 60% in part (a) and about 40 – 70% in part (b).

As an individual problem, the FR and MC results are not consistent and we see a high percentage of students are not able to correctly explain their reasoning, as evidenced by the small number of CC answers and high number of CI answers in the FR formats. The FR formats, however, do give us a chance to explore the types of CP, CI and I answers, in order to obtain information on students thought processes, critical thinking and problem solving skills. We discuss the types of incorrect answers in part IV.A.2.

A short explanation is needed in order to interpret the results of part (b) of the FR explicit ranking version given in Fall 2010. It is not possible to rank cases 2 and 3 in the problem, without making assumptions. So, without an explicit assumption and correct follow through, the answers were not counted as correct.

## 2. Solution types

We also studied the type of incorrect solutions given by the students. Graphs of the types of solutions for the CP, CI and I categories for the calculus-based course are given in Figures 3-5. In Figures 3 and 4, we plot the most common types of incorrect answers for the calculus-based course for problem A by problem format. In Figure 5, we plot the most common incorrect solutions for all problem formats together. As the same categories were observed in the algebra-based course, with similar distribution, we limit the examples to the calculus-based course.

## (a) CP and CI answers in part (a)

In part (a) of the problem, the majority of CP and CI answers fell into four main categories, labeled in the graphs (1) "height", (2) "same force", (3) "lacking detail", (4) "statement" and (5) "other." They are listed by number below:

1) Students addressed only the fact that the height difference was the same. They did not discuss conservation of energy. They simply stated that the final speed was the same in each case because the balls all started from the same height. There is no explicit evidence that the students understand the conservation of energy or understand the need to apply a consistent, logical argument, based on physics principles in an answer to a quiz question.

Examples of this type of answer are:

   (a) The speed is the same for all three cases since they start at the same height.

   (b) Same. Same starting height and same finishing line.

2) Students argued that the gravitational force was the same in each case or that they each had the same acceleration. They made statements about the gravitational force without any connections or reference to physics principles. They did not address the distance traveled, either vertically or in the direction of motion.

   Examples of this type of answer:

   (a) They all have the same speed, because they all have the same amount of gravity acting on them.

   (b) Their speeds will all equal because gravity has the same acceleration on all of them.

   (c) They have the same speed, since the only effect is mg since there is no friction.

3) Students discussed the conservation of energy in words or equations, but did not explicitly address the initial and final conditions or were lacking other important parts of a logical argument. The majority of CP answers fall into this category. In this case, it is possible that the students do understand the physics concepts required to answer the problem, but that the instructors do not usually require them to write out their reasoning, demonstrating clearly all of the steps and logic. This may also be related to the fact that the most of the course assessment (exams and online homework) was in MC format and students were rarely, if ever, required to write out their calculations and reasoning for grading.

   Examples of this type of answer are:

   (a) They would all be the same speed because $mgh = 1/2\ mv^2$.

   (b) Same speed. Mgh PE to KE. The PE to KE means that all of the speeds are the same.

4) Students in this category answer with a statement and do not show any explanation or calculations.

   An example of this type of answer is:

   They all reach the bottom at the same speed.

Almost all of the answers fall into one of the above categories. Other types of answers do not occur frequently enough to form a category of a common answer.

**(b) Incorrect (I) answers in part (a)**

The majority of Incorrect answers fell into five distinct categories, labeled in the graphs (1) "largest net force", (2) "longer distance", (3) "longer time", (4) "incorrect argument," (5) "centripetal" and (6) "other." They are listed by number below:

1) The students argued that the final velocity of the ball in case 1 is greatest, because it has the largest net force acting on it.

   Examples of this type of answer are:

   (a) Dropping the ball will make it reach the bottom with the highest speed because no other force is acting on it except gravity.

   (b) 1. because gravity is acting on the ball and it is falling at the max. possible speed assuming no air resistance. In the other problems, the ramp and the ball swinging have more forces acting on them.

2) The students argued that the ball in case three would arrive at the bottom with the highest speed because it travels a longer distance.

   Examples of this type of answer are:

   (a) It has the farthest to move therefore its speed can increase more.

   (b) Ball 3 will reach the bottom with the highest speed, because it travels the farthest and builds up the most speed.

3) The students argued that the ball in case three would arrive at the bottom with the highest speed because it travels for a longer time.

   (a) swinging takes a little longer than dropping, so it acquires more speed at the bottom. Then dropping because of gravity, then sliding would have the slowest velocity due to the incline of the ramp.

   (b) as the object falls longer, it accelerates more due to gravity, thus the longer the traveling of the object, the faster its speed.

4) Students answered with incorrect physics or illogical arguments.

   Examples of this type of answer are:

(a)    3 because ball 3 has g and a.

(b)    3 swinging down, because in case 1 it comes to a stop, case 2 slows down after the ramp, but in case 3 the ball swings through.

5) Students in this category invoked centripetal motion or centripetal force to explain the higher speed of the ball in Case 3.

Examples of this type of answer are:

(a) The ball swinging down has more than just gravity acting upon it. It has centripetal acceleration plus its mass times gravity.

(b) Yes, its centripetal motion will allow it to pick up speed faster than both 1 and 2.

(c) swinging down would have the highest speed, because it is acted on by gravity as well as centripetal force.

**(c) CP and CI answers in part (b)**

The majority of CP and CI answers to part (b) fell into three categories, labeled in the graphs as: (1) "shortest path", (2) "greatest net force", (3) "statement" and (4) "other." By far the most common was to state that the ball in case 1 arrived first, because it had the shortest path. These students did not address the acceleration as part of their answer. The second most common category was the students who answered that the ball in case 1 would arrive first because it had the greatest net acceleration on it. These students did not explicitly address the distance traveled. The categories are listed by number as:

1) Students answered that the ball in case (1) would arrive first because it took the shortest path or that it had the most direct route.

Examples of this type of answer are:

(a) 1 because it is falling straight down.

(b) #1 Most direct route.

2) Students answered that the ball in case (1) would arrive first because it had the greatest net force acting on it.

An example of this type of answer is:

Case 1 because there is nothing acting on the ball except gravity so it drops to the bottom first.

3) Students answered with a statement only.

An example of this type of response is:

Case 1. It goes directly to the bottom.

## (d) Incorrect (I) answers in part (b)

The majority of Incorrect (I) answers in part (b) fell into six categories, labeled in the graphs as (1) "same force", (2) "same height", (3) "statement", (4) "energy," (5) "same speed," and (6) "incorrect argument." They are listed by number below:

1) Students answered that all of the balls would reach the bottom at the same time because they had the same force of gravity acting on them.

Example of answers in this category are:

(a) since there is no friction acting on the ball in part 2 and all the balls are experiencing only the force of gravity.

(b) They will all reach the bottom at the same time because gravity is constant.

2) Students argued that the balls should reach the bottom at the same time because they fell through the same height.

Examples of answers in this category are:

(a) The balls should all reach the bottom at the same time, since they all start from the same height and no friction is involved.

(b) All 3 cases, objects fall at same height and time.

3) Students answered with just a statement.

An example of an answer in this category is:

They all reach the bottom at the same time.

4) Students argue that all of the balls will reach the bottom at the same time based on an energy argument.

Examples of the type of answers in this category are:

(a) They will all reach the bottom at the same time, because they all start off with the same potential energy.

(b) They all have potential energy and will be kinetic and it will be the same, so the time will be the same. They will all reach the bottom at the same time, for case 2 and 3 can be assumed to be like a free fall, since mass gravity and height are all the same.

5) Students argue that because the balls arrive with the same speed, they will arrive at the same time. (Some students argue that the balls are traveling with the same speed.)

An example of the type of answer in this category is:

(a) because they are travelling at the same speed at the bottom, they all hit the bottom at the same time as well.

6) Students make incorrect arguments, incorrectly apply physics concepts or do not make logical arguments:

An example of the type of answer in this category is:

Again same speed for all. We were shown in class that two objects (one shot horizontally and one dropped straight down) would hit the ground at the same time.

## B. Results for written responses for Problem B

After we had administered six versions of Problem A, we decided to change the problem used in the recitation quiz at that point in the course. We chose a different problem that also focused on the concept of conservation of energy and the understanding and use of kinematics equations and Newton's Second Law, depending on how one worked the problem. It was a problem similar to those in *Ranking Task Exercises in Physics: Student Edition.*[23] It was administered in three formats, MC/FR, FR and FR Calculate. The first two formats, MC/FR and FR, required an explicit ranking of four cases in the answer. This was different from Problem A. Answers to Problem A only required a choice or defense of a single scenario as the answer, except in the case when explicit ranking was written into the stem. Ranking was implicit in most versions of Problem A. The third version of Problem B, FR Calculate, did not ask for a ranking, but for a calculation of a quantity for each of the four scenarios.

The results were significantly different in this case. In Problem B part (a), we did not see the preponderance of CI answers that we saw in Problem A and in Problem B part (b), we have identified almost all of the CI answers as the incorrect use of a kinematics equation that gave the correct ranking, which is something that could be changed in future versions.

In short, the number of CI responses decreased dramatically when we administered a "Ranking Task" problem of the type found in *Ranking Task Exercises in Physics: Student Edition,*[23] a problem which required explicit ranking in the solution. The version asking for a straight calculation of a quantity for each case also did not have a significant percentage of CI answers.

As there were very few CP answers, we analyzed the solution types of the CI and I answers only in this case.

## 1. Numerical analysis

The results for both the calculus-based and algebra-based sections for Problem B, administered in Fall 2011, are shown in Tables 1 and 2 and Figure 6. The percentages of CI in part (a) are very low compared to those found in Problem A, with the most found in the MC/FR version, where it is possible to guess the correct answer. The higher percentage of CI in part (b), as mentioned above, can be explained and will be discussed below.

## 2. Solution types

The algebra-based and calculus-based answers were again similar, and so we have chosen this time to use the algebra-based results as an example. We graph and report on all of the common incorrect answer types (CI and I) for the algebra-based course for Problem B, parts (a) and (b) in Figure 7 and below.

### Incorrect solutions for part (a)

There are four main categories of incorrect answers (I and CI) in part (a). The categories are labeled in the graphs as (1) "incorrect conservation of energy," (2) "based on height," (3) "based on steepness/slope/angle," (4) "no indication of process," (5) "other," and (6) "correct." The incorrect types of answer are listed by number below:

1) Students had errors in their use of conservation of energy. One common error was to assign the initial potential energy the value PE = mghsinθ. There were also calculational errors in this category.

2) Students made arguments based on height, without referring to conservation of energy.

   An example of this is:

   Both situation B and D have a greater velocity than A and C because their heights are greater.

3) Students made arguments based on steepness, slope or angle.

   For example, one student showed work and correctly calculated the angle of the slope in each case. The student then ranked the final velocities B, C = D, A, based on the calculation of angles. The student then wrote: "The bigger angle provides a steeper slope, causing the velocity of the mass on the steepest slope to have the greatest velocity."

4) Students did not indicate any process at all and simply wrote down an answer.

**(a) Incorrect solutions for part (b)**

There are six common incorrect answers (I and CI) for part (b). The categories are labeled in the graphs as (1) "t =d/v," (2) "single kinematics eqn.," (3) "distance and slope or height," (4) "velocity and distance or height," (5) "angle," (6) "no indication of process," (7) "other," and (8) "correct." The incorrect types of answer are listed by number below:

1) The most common type of incorrect answer accounts for the students in the CI category. They simply wrote down the equation t = d/v, set d equal to the distance down the slope and v equal to the final velocity. This gives the correct ranking, but is not a correct solution to the problem. If the students had used the average velocity, instead of the final velocity, their process would have been correct. The students did not consider acceleration in the choice and application of the equation. In many instances, it appeared as if they simply chose the kinematics equation they did, because they had a length given in the problem and they had a velocity they had just calculated in part (a). In other words, they were not thinking about the physics of the problem, but how to get an answer.

2) Students tried to use a single kinematics equation, such as $v = v_0 + at$. As they did not have a value for the acceleration, they arbitrarily chose one, such as 9.8m/s^2. To correctly work the problem one needs either to use a kinematics equation that does not include the acceleration as a variable or to use two kinematics equations (or a force equation and a kinematics equation) to solve for the time, as the acceleration is also unknown.

3) Students based their ranking on the distance traveled and slope or height, giving an incorrect ranking. They would argue that B has the "highest combination of height and less length" and A has the lowest height and longest length". They did not distinguish between C and D or simply wrote down a ranking for C and D without making an argument.

4) Students made an argument based on velocity and distance or height.

An example of this kind of argument is:

A > C = D  > B.  A travels slowest over the longest distance C = D because D moves twice as fast over twice the distance and both reach the bottom at the same time.

5) Students made an argument based on the angle. This did not distinguish C and D, so most of these students ranked A > C = D > B or the reverse.

6) Students gave no indication of process.

## C. General Comments on Incorrect Answers

We have given examples of common CP, CI and I responses to Problems A and CI and I responses to Problem B. We would like to note that we were also struck by the sparseness of the wording of many of the answers and lack of coherent and logical reasoning based on physics principles. Many students who answered the problem incorrectly focused on the surface features of the problem and not the basic physics principles involved. This is consistent with PER results in research on the problem solving methods of traditionally taught students. For example, in Problem B, students who argued any of the answers based on height, slope or angle, without the use of conservation of energy or a correct kinematics equation, were not basing their arguments on physics principles, but on other cognitive resources,[28] such as phenomenological primitives (p-prims),[29] also consistent with PER research. The types of answers revealed knowledge of students thought processes and incorrect conceptions that are not available to instructors simply looking at the numerical scores of MC exam results.

## D. Problem Format Survey

The results of the Problem Format Survey (Appendix III) are presented in Figure 8. The students were asked to identify 1) the type of physics exam or quiz question they prefer to be graded on, 2) the type of physics exam or quiz question they think is easiest and 3) the type of physics exam or quiz question they think gives their instructor the most information on their understanding of the question being asked. They were to choose one of five choices and explain their reasoning. The choices were: a) MC, b) true/false (T/F), c) short answer, d) show your work (including calculations) or explain your reasoning and e) Other (describe).

While MC has the highest percentage as the type of problem that students prefer to be graded on, it is actually not the majority of students when one considers that the students who chose the Other category primarily wanted a mixture of formats, including some which required show your work and explain your reasoning. If one considers Short Answer, Show Work/Explain and Other together, that is a higher percentage (61%) than Multiple-choice.

Students who chose either MC format or FR formats have common supporting arguments for their choice. Students choosing MC do so 1) because you can guess and 2) because it gives you an idea of where to start, if you are clueless. The main argument for preferring FR responses that require showing your work, calculation, explaining your reasoning or short answer are that you get partial credit.

MC and T/F together (74%) are considered by students to constitute the easiest type of questions, which has a lot to do with guessing options.

Show your work/Explain your reasoning combined with Short Answer and Other, all of which require written explanations, constitute (95%) the type of questions that students perceive give instructors the most information on their understanding. It is clear that the students perceive these question formats as harder, but that they feel they much better reflect their understanding of the content and skills being assessed, compared to MC and "other".

## V. DISCUSSION AND CONCLUSIONS

We have presented data from a large university where, in the large introductory classes, MC is the dominant form of exam assessment and most of the lecture instruction is in traditional lecture format. The majority of the homework is online and the students are usually not required to demonstrate an understanding of their thinking or solution process either in words or equations.

In Problem A, we observed a high percentage of CI answers and the inability of students to provide a consistent, coherent answer in either words or equations. Students with incorrect answers did not use physics principles to construct an answer, but arguments based on surface features of the problem. This information was not available to instructors based on the MC exams or the online homework. It was only available when FR questions were introduced in the recitation sections. If one goal of the courses was to be able to solve problems and answer questions using physics principles and consistent, coherent logic, a high percentage of the students failed, even though approximately 70% of the students passed the course with an A, B or C based on MC exam scores. The FR testing yielded results on students' thinking processes that were not available from the MC testing; it yielded results crucial as an information source to inform instructors about their students thinking skills and for use in informing future instructional methods and materials. While one can argue that this is just a single problem and it is possible to create an exam of mostly or only multiple choice questions that distinguishes A, B and C students,[7,28] a study of this kind suggests that the students may not be reaching answers by consistent, coherent logical arguments, may not be using physics principles to get to an answer and may be relying only on surface features of the problem or even "test-taking strategies" to get the right answer. Problem A has demonstrated how prevalent this was at one large university. We have seen similar results on many FR quizzes in our introductory courses over the course of our large scale assessment.

In problem A, we also see that if the question had been asked as a MC question, a high percentage (over 50% on each part of the problem) of the students would have chosen the correct answer. However, if asked to explain their answer, either with or without the benefit of choices, the majority of students would not be able to do it correctly. Unless one is administering an exam that contains only problems that have been sufficiently well tested to eliminate this high percentage of false positives, this effect may occur fairly often on exam problems. While a PER group or someone interested in PER may develop MC exams that decrease the number of false positives, the average instructor at a large university may not always have or take the time to do this.

One can argue that the ability to demonstrate the process of obtaining a correct answer through physics principles, is a skill as important as being able to choose an answer that illustrates a correct conceptual understanding. We have demonstrated that when thinking process and critical thinking skills are not explicitly assessed (which was the case in large lecture classes in this study assessed primarily by MC exam questions), there is no evidence that students have achieved this skill. Indeed, when we did assess process skills through FR assessment, we found that many students did not have these skills.

In Problem B, we see evidence that a problem requiring explicit ranking, and possibly the use of more MC options, might decrease the number of false positives in MC questions. The false positives in the (b) part of that problem could easily have been eliminated, if the incorrect use of a kinematics equation that gave the correct ranking had been identified before the question was administered. We could argue that the explicit ranking added a higher order of complexity, and that complexity, coupled with the increased number of MC options, decreased the number of false positives compared to Problem A (which simply required the choice of one of four scenarios as a correct answer). This is evidence of one way that MC questions could be constructed to decrease the number of false positives.

However, the main point is that in large lecture classes at a large university, where approximately 70% of the students finish the course with an A, B or C, up to 90% of the students could not choose a correct answer <u>and</u> correctly explain their reasoning or show their work on quiz problems such as the two sample problems in this paper. Yet, they did sufficiently well on MC format exams in lecture and online homework to pass the course. If the goal of the course is simply to distinguish A, B and C students based on their numerical scores on MC format exams, without understanding that a high percentage of the students were not able to get to the correct answer by a logical, consistent application of physics principles, but by some ad hoc methods, then the MC format is sufficient. However, if the goals of the course include developing process and critical thinking skills in students, assessed by the demonstration of those skills on assessment instruments, then the MC format exams have failed. In addition, the instructors of these courses who have only used the MC assessment format have no information on the types of incorrect answers and incorrect reasoning used by their students.

While we have only used two problems as examples, we have seen this multiple times in other problems. We suggest that more research be done on the skills assessed by MC format compared to the skills assessed by FR format, because the skills assessed must match the goals of the course and students must be able to demonstrate these skills.

The students' response to the Problem Format Survey, taught us that students believe that free response questions give instructors a much better understanding of the students' understanding of the concepts and their ability to demonstrate this through the use of calculation and explanation. We also found that the students think multiple-choice and true/false are easier, partly because you can guess, but also because it "gives you a clue," if you don't have any idea how to work the problem. While multiple-choice questions can give you ideas to start with, many students also are appreciative of the fact that with free response questions, they get partial credit for demonstrating that they do know some of the concepts and how to apply them.

Further research also needs to be done on how the students study habits are informed by the instructor's choice of exam format and how it promotes or doesn't promote the development of the desired skills.

In summary, we have evidence that 1) up to 90% of the students could not choose a correct answer <u>and</u> correctly explain their reasoning or show their work on the two typical quiz problems presented, 2) there were a high percentage of CI answers based on the FR and FR/MC results, 3) many students do not choose MC answers based on physics principles, but on surface features, and 4) FR format results can be inconsistent with MC results, as demonstrated at one a large university where traditional teaching, MC exam format and online homework are the norm. We also have evidence that "Ranking Task" problems, as well as an increased number of MC options, may decrease the number of false positives in MC format problems.

We suggest that the goals of a physics course should include the ability to demonstrate the process of obtaining a correct answer in words or equations through physics principles and that, if this is a goal, the ability to do this needs to be assessed. We believe that this goal is at least implicit in most physics courses and that it is explicit in many non-traditional and PER-informed courses. We have presented data that demonstrate that it cannot be assumed that students have these skills simply based on MC exam results and that FR or FR/MC formats better assess these skills. We suggest that assessment instruments used in large classes at large universities and instruments used in comparing students taught in different instructional formats, such as traditional and non-traditional, should include question formats that explicitly assess the thinking process and use of physics principles in solving problems, if those are some of the goals of the courses.

## VI. ACKNOWLEDGEMENTS

We are grateful to the National Institutes of Health (NIH) for the funding that made this study possible. We also thank Hani Dulli for help in the choice of one of the problems.

This article is based upon work supported by the National Institutes of Health (NIH) under Grant NIH 5RC1GM090897-02. Any opinions, findings and conclusions or recommendations expressed are those of the authors and do not necessarily reflect the views of the NIH.

| Assessment Format | CC | CP | CI | I | | CC | CP | CI | I |
|---|---|---|---|---|---|---|---|---|---|
| Fa 10 A (a) | | | | | Fa 10 A (b) | | | | |
| MC | 69 | 0 | 0 | 31 | | 55 | 0 | 0 | 45 |
| FR | 14 | 7 | 33 | 46 | | 0 | 4 | 62 | 34 |
| FR Exp. Rank | 17 | 13 | 25 | 45 | | 0 | 0 | 1 | 99 |
| Sp 11 A (a) | | | | | Sp 11 A (b) | | | | |
| MC/FR | 10 | 10 | 42 | 37 | | 1 | 3 | 68 | 28 |
| FR Format (e) | 10 | 5 | 61 | 24 | | 0 | 2 | 57 | 40 |
| FR Format (f) | 5 | 5 | 47 | 42 | | 1 | 4 | 73 | 23 |
| Fa 11 B (a) | | | | | Fa 11 B (b) | | | | |
| FR Calculate | 35 | 0 | 18 | 64 | | 4 | 0 | 7 | 89 |
| MC/FR | 25 | 0 | 15 | 60 | | 5 | 0 | 60 | 35 |
| FR | 21 | 1 | 4 | 74 | | 7 | 0 | 41 | 51 |

Table 1. The percentage of students in each answer category by assessment format for the calculus-based course. The data is listed by semester and problem part, for example, Fa 10 A (a) refers to Fall 2010 Problem A, part (a). The solution categories are I (Incorrect), CI (correct choice, incorrect explanation), CP (correct choice, partially correct explanation) and CC (correct choice, correct explanation).

| Assessment Format | CC | CP | CI | I | | CC | CP | CI | I |
|---|---|---|---|---|---|---|---|---|---|
| Fa 10 A (a) | | | | | Fa 10 A (b) | | | | |
| MC | 55 | 0 | 0 | 45 | | 52 | 0 | 0 | 48 |
| FR | 7 | 7 | 14 | 71 | | 1 | 1 | 70 | 27 |
| FR Exp. Rank. | 9 | 2 | 28 | 62 | | 0 | 0 | 0 | 100 |
| Sp 11 A (a) | | | | | Sp 11 A (b) | | | | |
| MC/FR | 2 | 9 | 42 | 47 | | 0 | 0 | 64 | 36 |
| FR Format (e) | 5 | 2 | 64 | 29 | | 5 | 0 | 42 | 53 |
| FR Format (f) | 7 | 2 | 40 | 51 | | 1 | 1 | 75 | 22 |
| Fa 11 B (a) | | | | | Fa 11 B (b) | | | | |
| FR Calculate | 48 | 0 | 0 | 52 | | 5 | 0 | 5 | 89 |
| MC/FR | 35 | 6 | 10 | 49 | | 0 | 0 | 43 | 57 |
| FR | 39 | 0 | 17 | 59 | | 0 | 0 | 37 | 63 |

Table 2. The percentage of students in each answer category by assessment format for the algebra-based course. The data is listed by semester and problem part, for example, Fa 10 A (a) refers to Fall 2010 Problem A, part (a). The solution categories are I (Incorrect), CI (correct choice, incorrect explanation), CP (correct choice, partially correct explanation) and CC (correct choice, correct explanation).

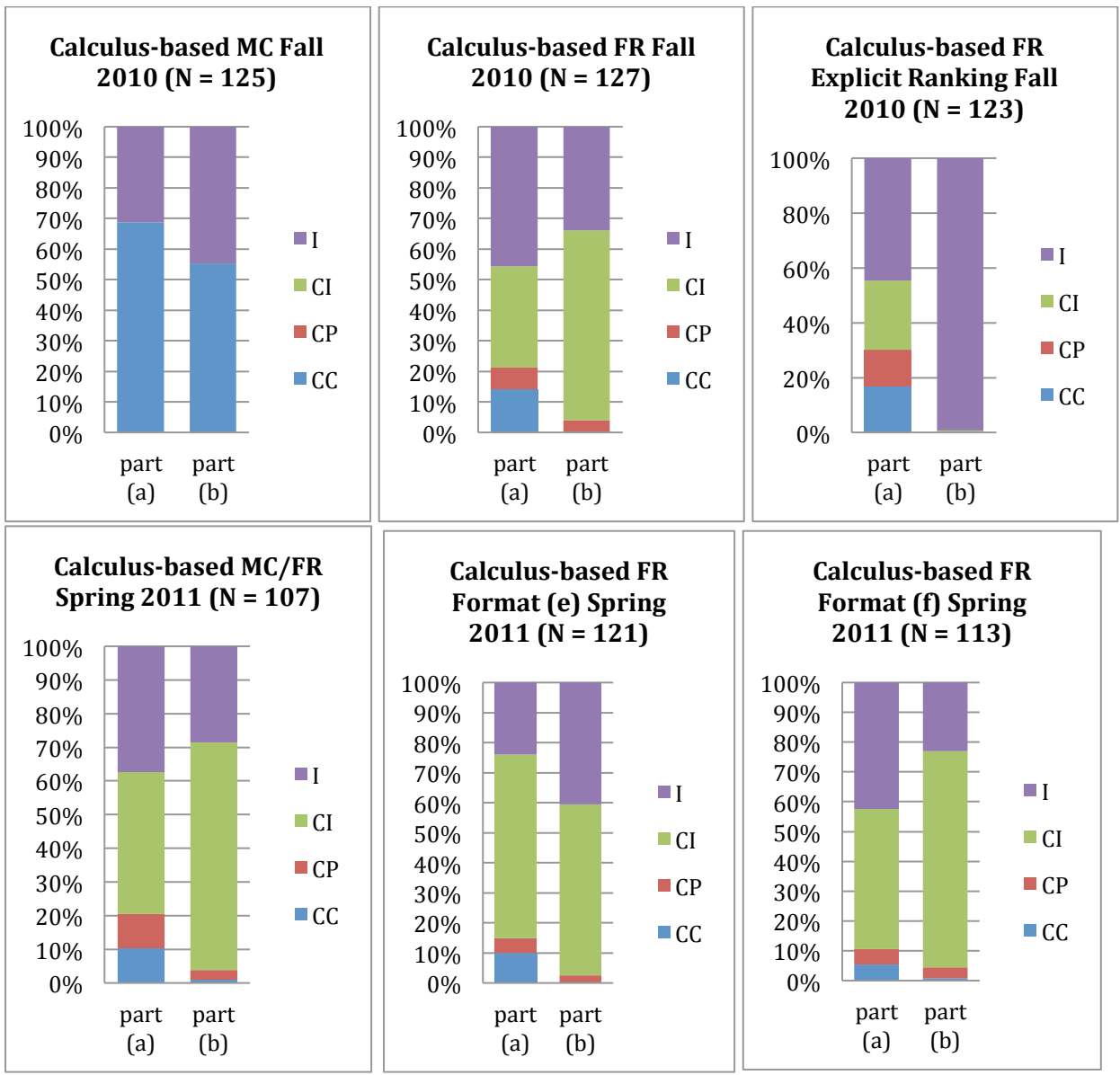

Figure 1: Calculus-based Results for Problem A. The categories are I (Incorrect), CI (correct choice, incorrect explanation), CP (correct choice, partially correct explanation) and CC (correct choice, correct explanation).

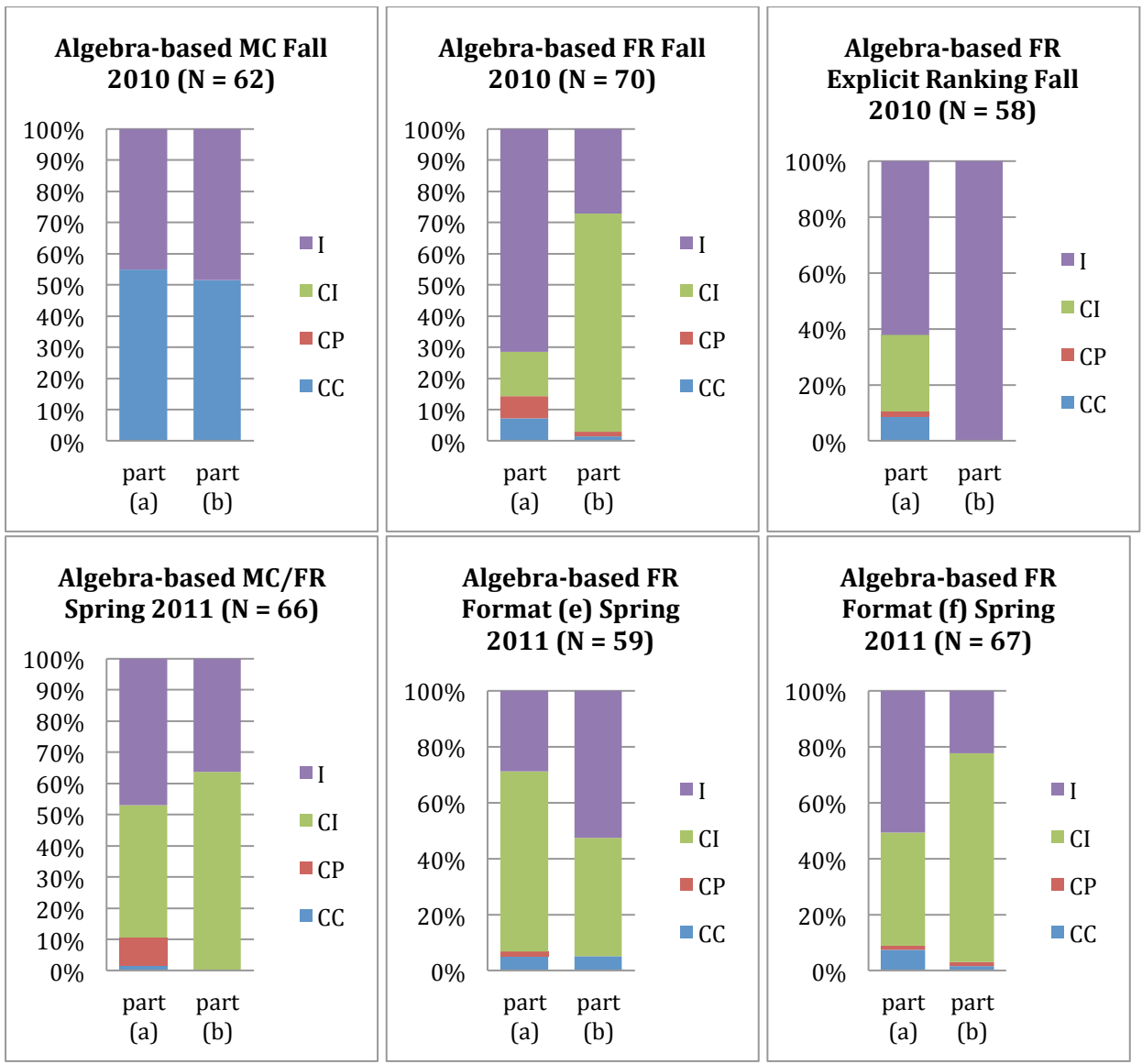

Figure 2: Algebra-based Results for Problem A. The categories are I (Incorrect), CI (correct choice, incorrect explanation), CP (correct choice, partially correct explanation) and CC (correct choice, correct explanation).

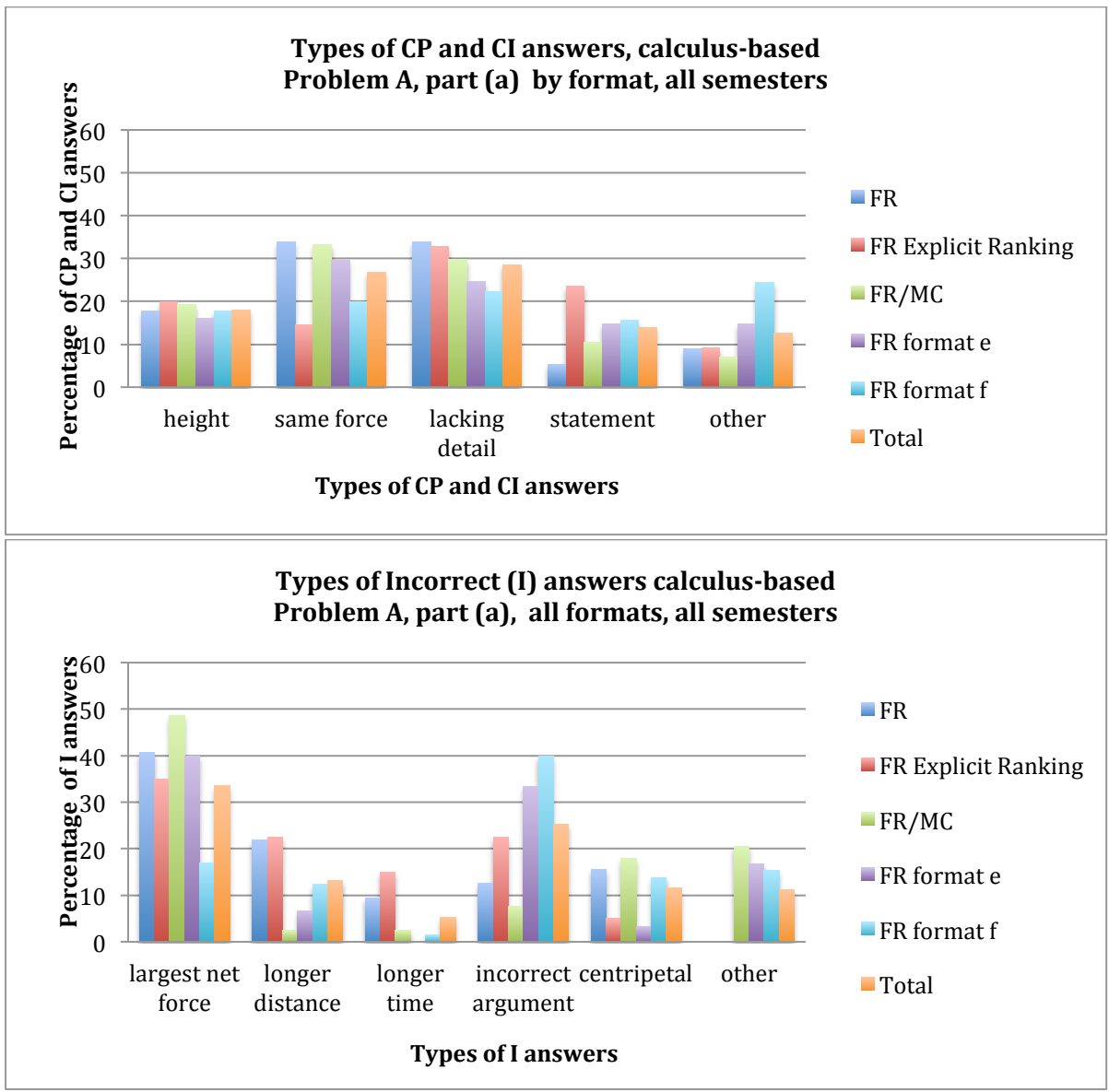

Figure 3: Percent of CP and CI answers by category of incorrect answer and problem format for calculus-based Problem A (a). The majority of CP and CI answers in part (a) fell into four main categories, labeled (1) "height", (2) "same force", (3) "lacking detail", (4) "statement" and (5) "other." The majority of I answers fell into five main categories, labeled in the graphs (1) "largest net force", (2) "longer distance", (3) "longer time", (4) "incorrect argument," (5) "centripetal" and (6) "other."

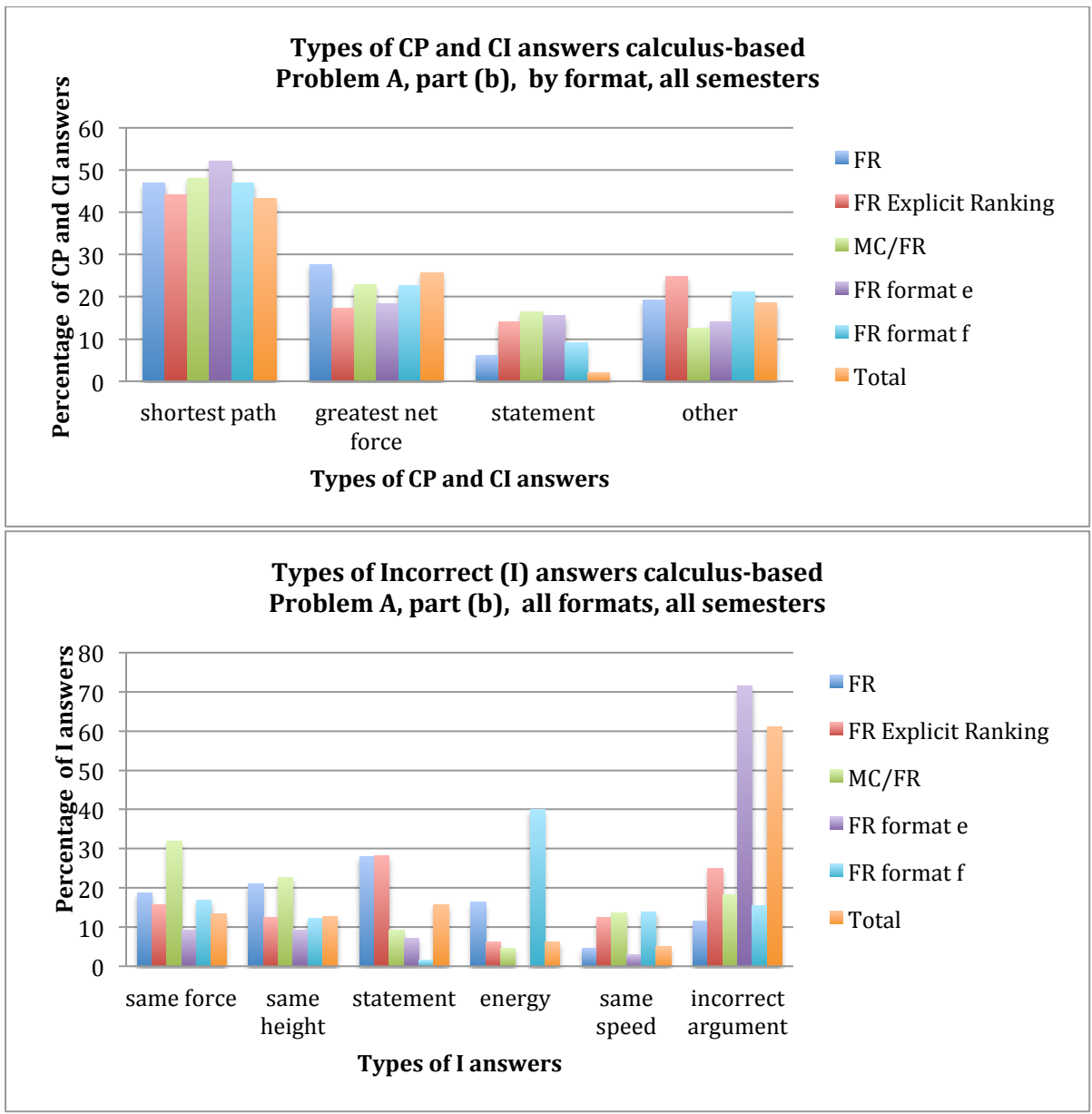

Figure 4: Percent of CP and CI and I answers by category of incorrect answer and problem format for calculus-based Problem A (b). The majority of CP and CI answers to part (b) fell into three categories, labeled in the graphs as: (1) "shortest path", (2) "greatest net force", (3) "statement" and (4) "other." The majority of I answers in part (b) fall into six categories, labeled in the graphs as (1) "same force", (2) "same height", (3) "statement", (4) "energy," (5) "same speed," and (6) "incorrect argument."

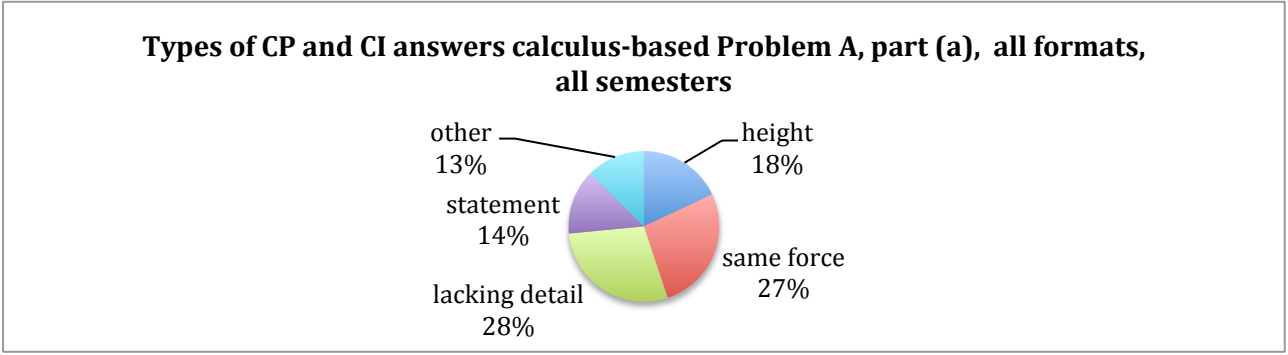

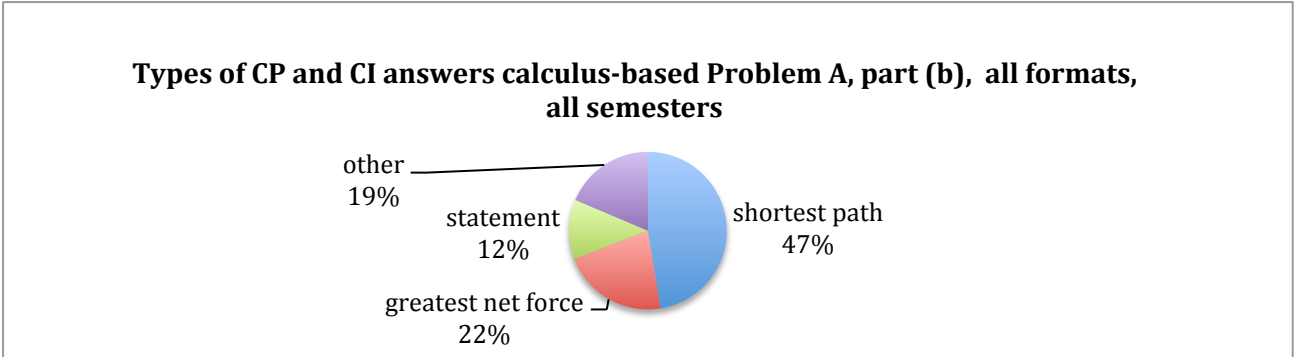

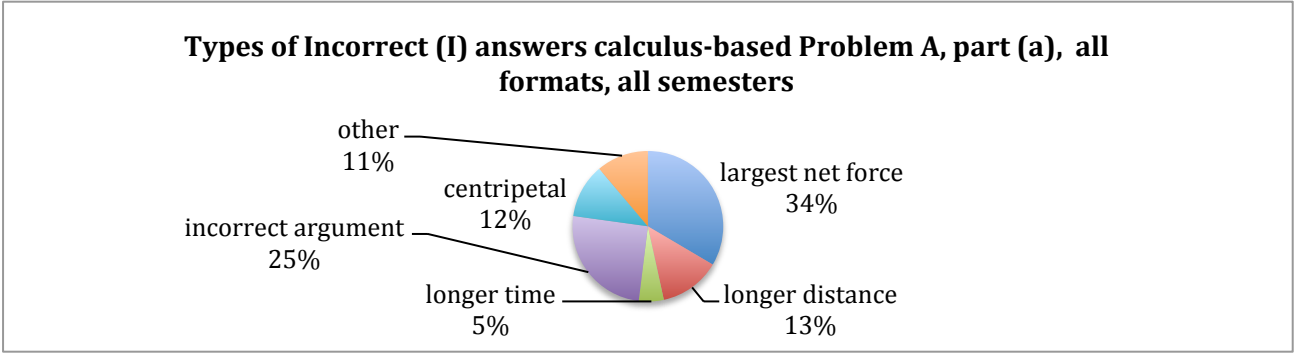

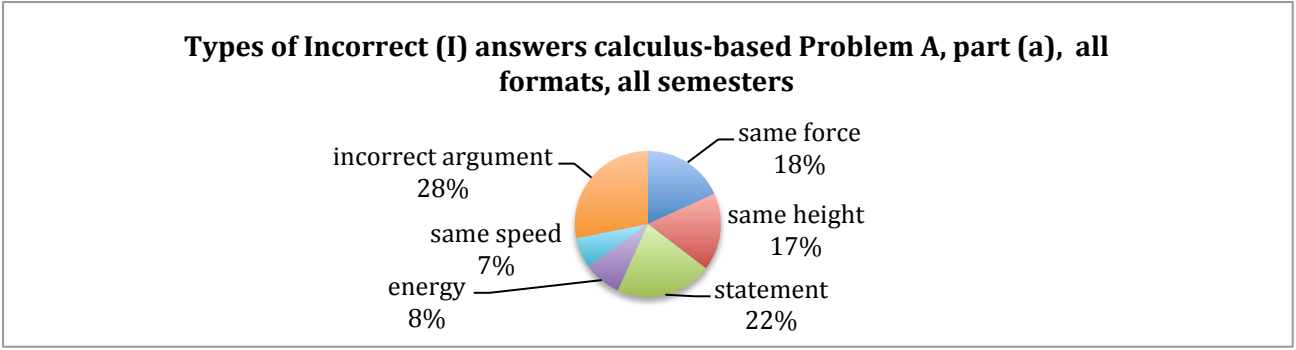

Figure 5: Types of incorrect solutions for the calculus-based course for Problem A for all format types for CP, CI and I answer types. The categories are the same as in Figures 3 and 4.

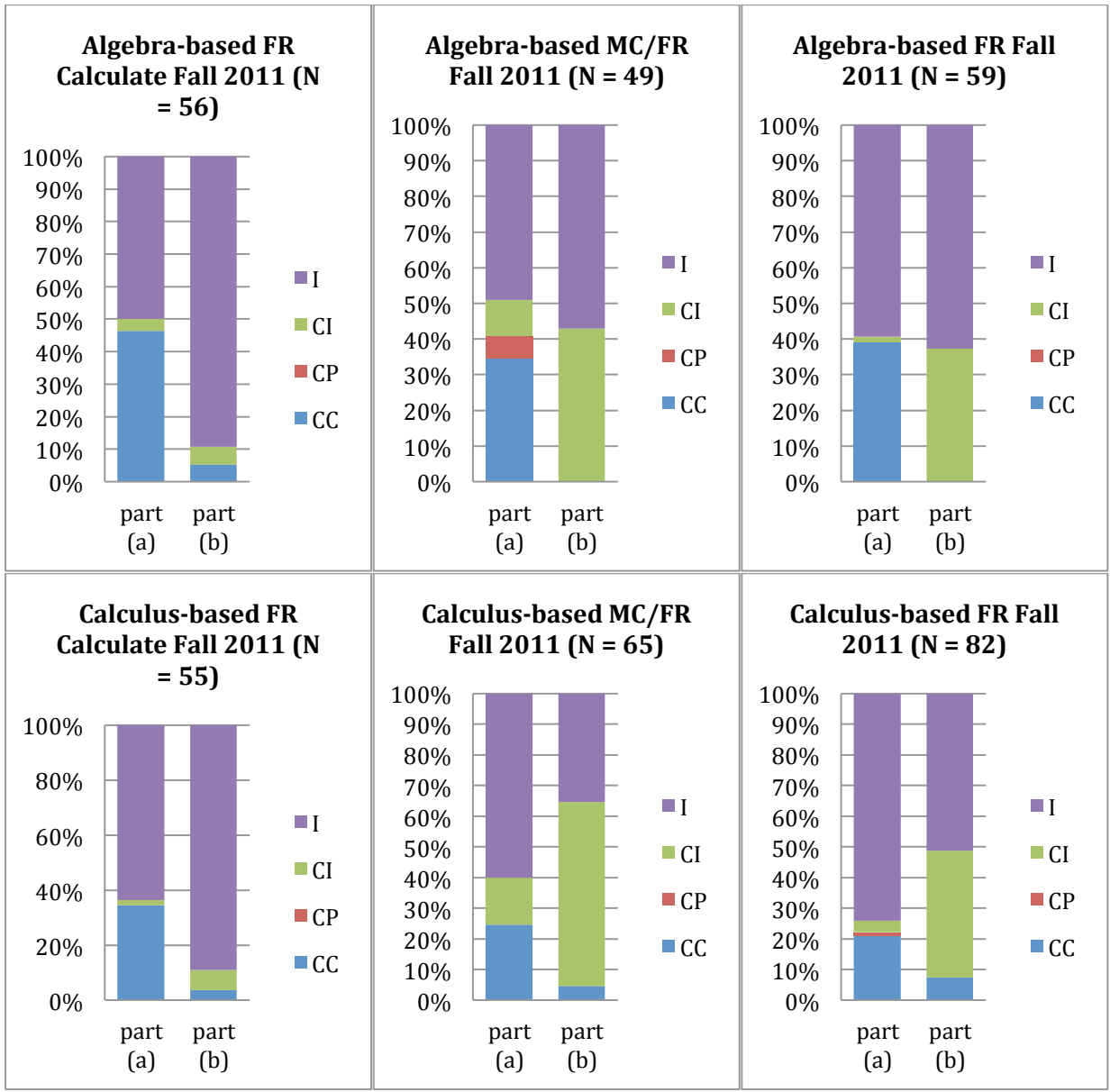

Figure 6: Algebra-based and calculus-based results for Problem B. The categories are I (Incorrect), CI (correct choice, incorrect explanation), CP (correct choice, partially correct explanation) and CC (correct choice, correct explanation).

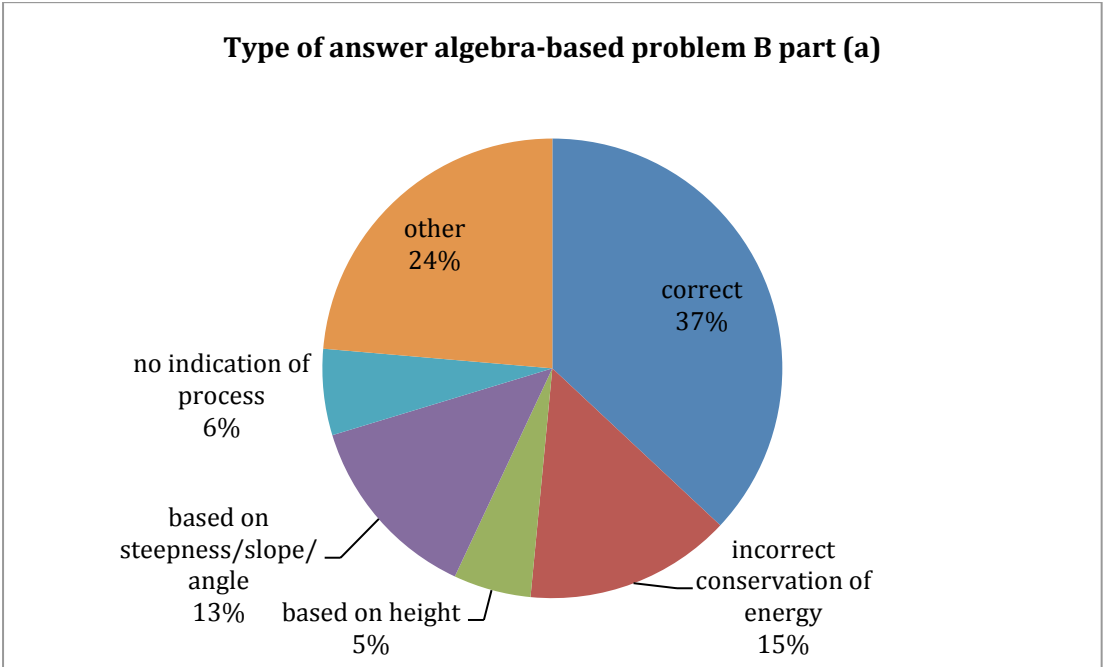

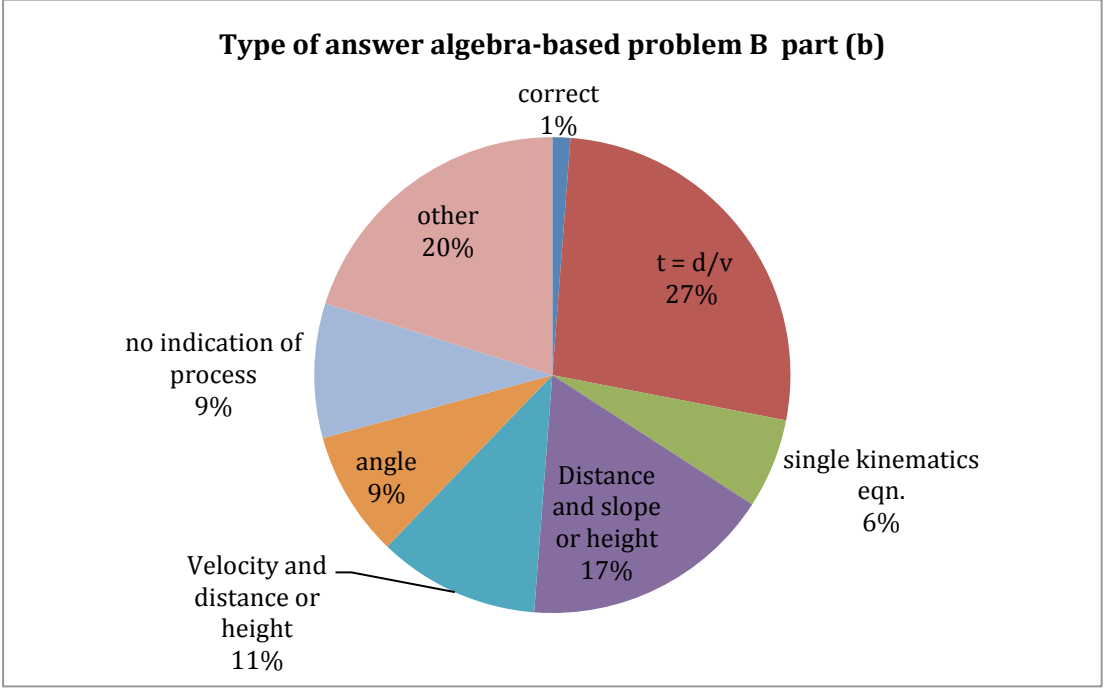

Figure 7: Distribution of common types of incorrect solutions for the algebra-based course for Problem B for all answer types. The categories in part (a) are labeled as (1) "incorrect conservation of energy," (2) "based on height," (3) "based on steepness/slope/angle," (4) "no indication of process," (5) "other," and (6) "correct." The categories in part (b) are labeled in the graphs as (1) "t =d/v," (2) "single kinematics eqn.," (3) "distance and slope or height," (4) "velocity and distance or height," (5) "angle," (6) "no indication of process," (7) "other," and (8) "correct."

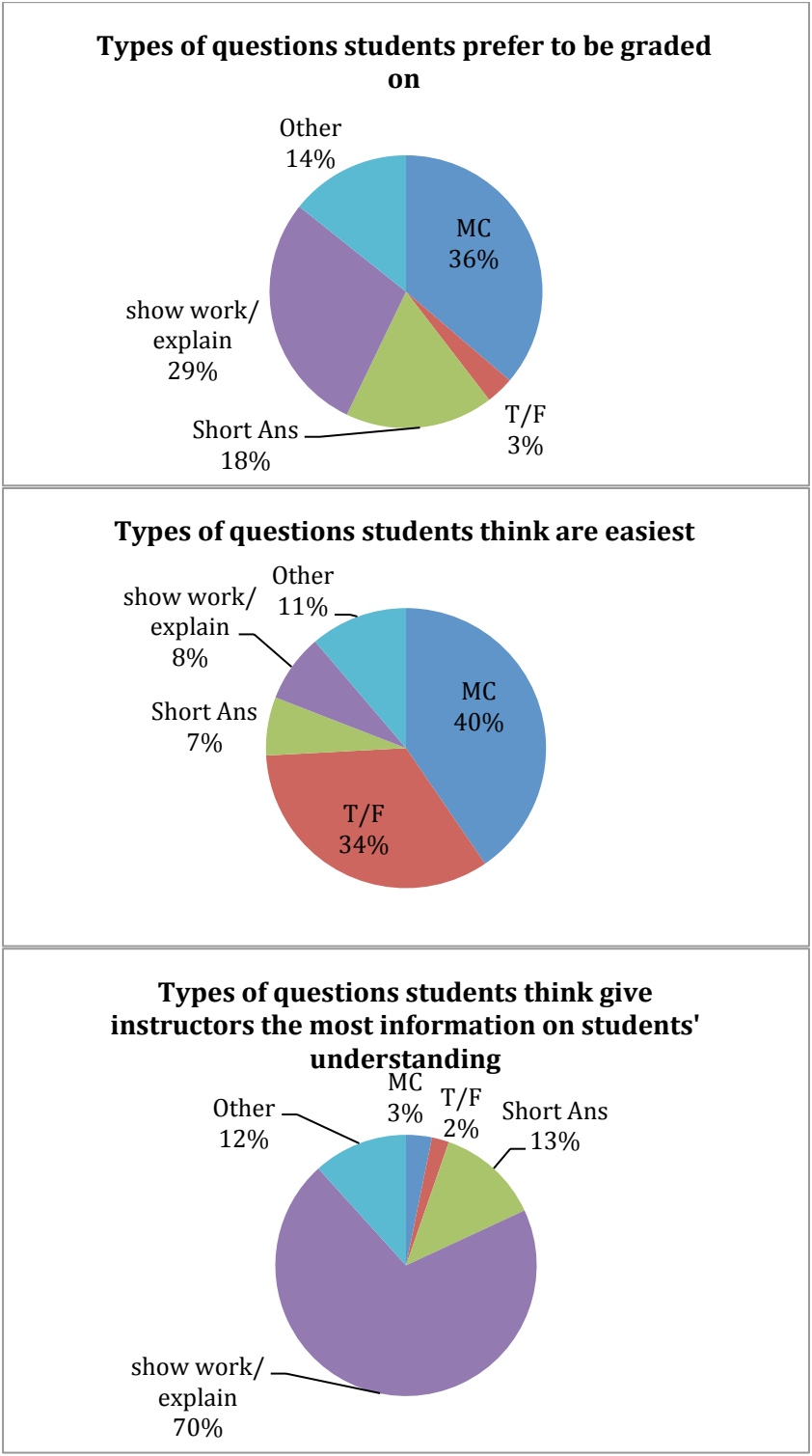

Figure 8: Student responses to the Problem Format Survey. Question type categories are: MC (multiple-choice), T/F (true/false), Short Ans. (short answer), show work/explain (show your work and/or explain your reasoning) and Other.

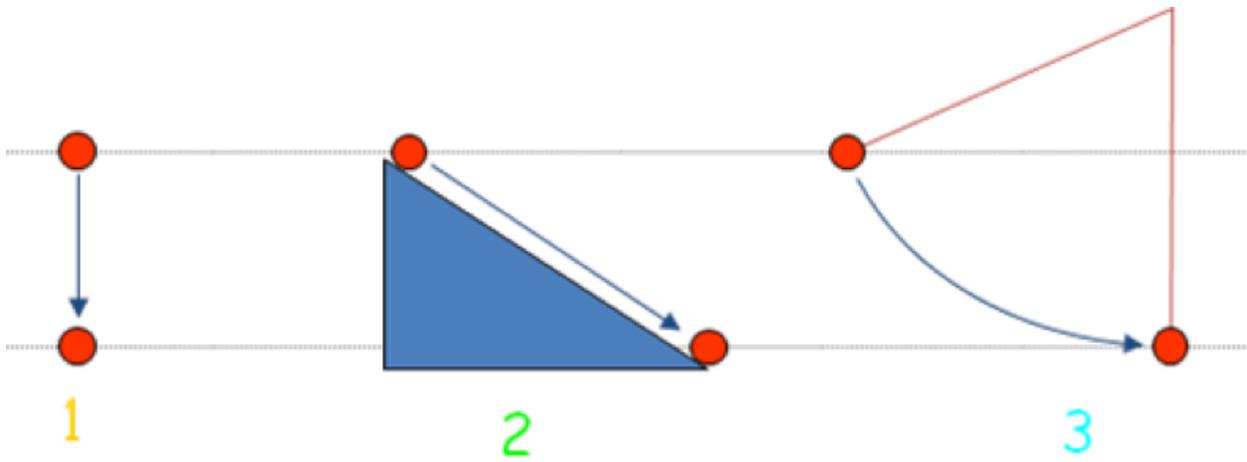

Figure 9: The picture referred to in the basic stem of Problem A in Appendix I.

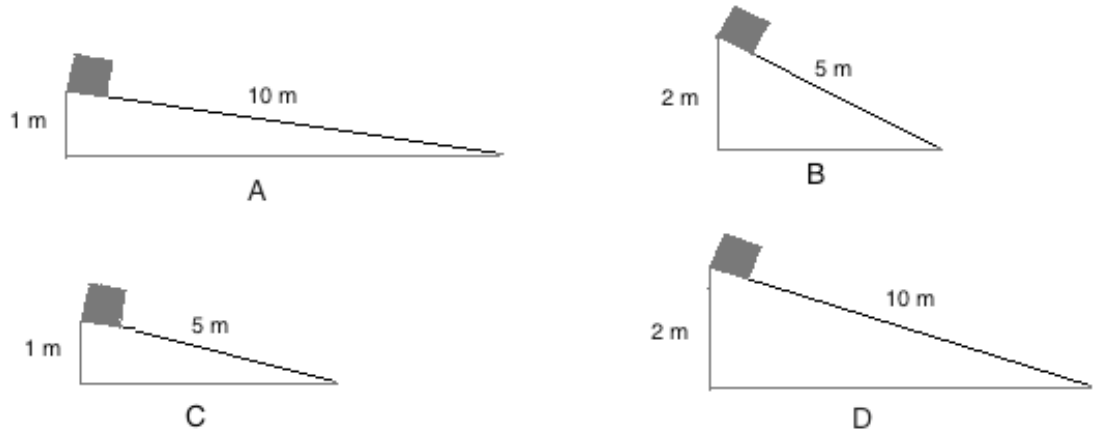

Figure 10: The figures referred to in the basic stem of Problem B in Appendix II.

# References

1. William C. Ward, "A Comparison of Free-Response and Multiple-Choice Forms of Verbal Aptitude Tests," Applied Psychological Measurement, January 1982; vol. 6, 1: pp. 1-11.

2. Brent Bridgeman , "A Comparison of Quantitative Questions in Open-Ended and Multiple-Choice Formats," Journal of Educational Measurement
   a. Volume 29, Issue 3, pages 253–271, September 1992.

3. David Thissen, Howard Wainer and Xiang-Bo Wang, "Are Tests Comprising Both Multiple-Choice and Free-Response Items Necessarily Less Unidimensional than Multiple-Choice Tests? An Analysis of Two Tests," Journal of Educational Measurement , Vol. 31, No. 2, Summer, 1994, 113-123.

4. William B. Walstad and William E. Becker, "Achievement Differences on Multiple-Choice and Essay Tests in Economics,' The American Economic Review, Vol. 84, No. 2, May, 1994, 193-196.

5. Gregory R. Hancock, "Cognitive Complexity and the Comparability of Multiple-Choice and Constructed-Response Test Formats," The Journal of Experimental Education,  Vol. 62, No. 2, Winter, 1994, 143-157.

6. Robert Lukhele, David Thissen, Howard Wainer, "On the Relative Value of Multiple-Choice, Constructed Response, and Examinee-Selected Items on Two Achievement Tests," Journal of Educational Measurement, Volume 31, Issue 3,  September 1994, pages 234–250.

7. Michael L. Scott, Tim Stelzer, Gary E. Gladding, "Evaluating Multiple-choice Exams in Large Introductory Physics Courses," Phys. Rev. ST Phys. Educ. Res. 2, 020102, (2006).

8. Michael E. Martinez , "A Comparison of Multiple-Choice and Constructed Figural Response Items," Journal of Educational Measurement, Volume 28, Issue 2,  June 1991, pages 131–145.

9. D. I. Newbie, Avril Baxter, R.G. Elmslie, "A comparison of multiple-choice tests and free-response tests in examinations of clinical competence," Medical Education, Volume 13, Issue 4,  July 1979, pages 263–268.

10. J.J. Veloski, H.K. Rabinowitz, M. R. Robeson, "A solution to the cueing effects of multiple choice questions: the Un-Q format," Medical Education, Volume 27, Issue 4, July 1993, pages 371–375.

11. Craig A. Berg, Philip Smith , "Assessing students' abilities to construct and interpret line graphs: Disparities between multiple-choice and free-response instruments," Science

Education, Volume 78, Issue 6, November 1994, pages 527–554.

12. Robert B. Frary, "Multiple-choice versus Free-response: A Simulaiton Study," Journal of Educational Measurement, Volume 22, Issue 1, March 1985, pages 21–31.

13. Veloski, J J; Rabinowitz, H K; Robeson, M R; Young, P R, "Patients don't present with five choices: an alternative to multiple-choice tests in assessing physicians' competence, Academic Medicine, Volume 74 - Issue 5, May 1999, pp. 539 – 536.

14. Bruce A Fenderson, Ivan Damjanov, Mary R Robeson, J.Jon Veloski, Emanuel Rubin, "The virtues of extended matching and uncued tests as alternatives to multiple choice questions," Human Pathology, Volume 28, Issue 5, May 1997, Pages 526–532.

15. R. N. Steinberg and Mel S. Sabella, "Performance on multiple-choice diagnostics and complementary exam problems," The Physics Teacher **35**, 150 (1997).

16. Robert J. Dufresne, William J. Leonard, and William J. Gerace, "Marking sense of students' answers to multiple-choice questions," The Physics Teacher **40** (3), 174, 2002.

17. N. Sanjay Rebello and Dean A. Zollman, "The effect of distracters on student performance on the force concept inventory," Am. J. Phys. **72** (1), 116-125, (2004).

18. National Institutes of Health (NIH) Challenge grant #1RC1GM090897-01, "An Assessment of Multimodal Physics Lab Intervention Efficacy in STEM Education," to assess for interventions to the laboratory curriculum at Texas Tech University, PI's Beth Thacker and Kelvin Cheng.

19. I. Halloun and D. Hestenes, "The Initial Knowledge State of College Physics Students," *Am. J. Phys.* **53**, 1043-1055 (1985).

20. In particular, we used the Colorado Learning Attitudes about Science Survey (CLASS), W. K. Adams, K. K. Perkins, N. S. Podolefsky, M. Dubson, N. D. Finkelstein, and C. E. Wieman, "New instrument for measuring student beliefs about physics and learning physics: The Colorado Learning Attitudes about Science Survey", Phys. Rev. ST Phys. Educ. Res. 2, 010101 (2006) and The Classroom Test of Scientific Reasoning (CTSR), Anton E. Lawson, "The development and validation of a classroom test of formal reasoning," J. Res. Sci. Teach. 15, 11 (1978).

21. We used the Reformed Teaching Observation Protocol (RTOP), MacIsaac, D.L. & Falconer, K.A. (2002, November). Reforming physics education via RTOP. The Physics Teacher 40(8), 479-485. for this purpose.

22. University of Illinois Physics Education Research website, http://research.physics.illinois.edu/per/, (08/16/12).

23. T. L. O'Kuma, D. P. Maloney and C J Hieggelke , *Ranking Task Exercises in Physics: Student Edition*, Benjamin Cummings, (Nov., 2003).

24. Katrien Struyven, Filip Dochy & Steven Janssens, "Students' perceptions about evaluation and assessment in higher education: a review," Assessment & Evaluation in Higher Education, Volume 30, Issue 4, 2005, pp/ 325-341.

25. Moshe Zeidner, "Essay versus Multiple-Choice Type Classroom Exams: The Student's Perspective," The Journal of Educational Research, Vol. 80, No. 6, pp. 352-358.

26. Karen Scouller, "The influence of assessment method on students' learning approaches: Multiple choice question examination versus assignment essay" Higher Education, June 1998, Volume 35, Issue 4, pp 453-472.

27. William L. Kuechler, Mark G. Simkin, "Why Is Performance on Multiple-Choice Tests and Constructed-Response Tests Not More Closely Related? Theory and an Empirical Test" Decision Sciences Journal of Innovative Education, Volume 8, Issue 1, January 2010, pages 55–73.

28. David Hammer, "Student resources for learning introductory physics," American Journal of Physics [0002-9505] **68** S52 (2000).

29. Andrea A. diSessa, "Toward an Epistemology of Physics," Cognition and Instruction, **10**, 105 (1993).

30. Shih-Yin Lin and Chandralekha Singh, "Can Multiple Choice Questions Simulate Free-Response Questions?" Physics Education Research Conference 2011, Omaha, Nebraska: August 3-4, 2011, **1413**, 47 (2011).

# Appendix I

## A. Problem A

**Basic stem:**

1) Imagine that you are comparing three different ways of having a ball move down through the same height, as in the picture below.

   [See Figure 9]

   The three cases are:
   1. Dropping
   2. Sliding on a ramp (no friction)
   3. Swinging down

**Version a (MC):**

a) In which case does the ball reach the bottom with the highest speed? Circle the correct answer.

1. Dropping
2. Slide on ramp (no friction)
3. Swinging down
4. All the same

b) In which case does the ball get to the bottom first? Circle the correct answer.

1. Dropping
2. Slide on ramp (no friction)
3. Swinging down
4. All the same

**Version b (FR):**

a) In which case does the ball reach the bottom with the highest speed? Explain your reasoning.

b) In which case does the ball get to the bottom first? Explain your reasoning.

**Version c (FR Explicit Ranking):**

a)  In which case does the ball reach the bottom with the highest speed? Rank the cases by the speed of the ball at the bottom. Show your ranking. Explain why you ranked them the way you did.  If it is not possible to rank the cases, explain why it is not possible.

b)  In which case does the ball get to the bottom first? Rank the cases in order of the time they reach the bottom from shortest to longest. Show your ranking. Explain why you ranked them the way you did.  If it is not possible to rank the cases, explain why it is not possible.

**Version d (MC/FR):**

a)  In which case does the ball reach the bottom with the highest speed? Circle the correct answer and explain your reasoning.

1.  Dropping
2.  Slide on ramp (no friction)
3.  Swinging down
4.  All the same

b)  In which case does the ball get to the bottom first? Circle the correct answer and explain your reasoning.

1.  Dropping
2.  Slide on ramp (no friction)
3.  Swinging down
4.  All the same

**FR Version e:**

a)  Is it possible for the balls in cases 2 and 3 to reach the bottom at the same speed as the ball in case 1? Explain your reasoning.

b)  Is it possible for the balls in cases 2 and 3 to reach the bottom at the same time as the ball in case 1? Explain your reasoning.

**FR Version f:**

a)  Is it possible for the ball in case 3 to reach the bottom with the highest speed? Explain your reasoning.

b)  Is it possible for the ball in case 3 to reach the bottom first? Explain your reasoning.

## B. Problem A Solution

A possible correct solution to Problem A (version b, Explain):

a) All balls reach the bottom with the same speed. They all start from the same height. Because, in each case, the gravitational force is the only force doing work, we can apply the conservation of energy, $mgh_i + 1/2mv_i^2 = mgh_f + 1/2mv_f^2$ to each of the cases. The initial and final heights and the initial velocity is the same each case. Therefore the final velocity, $v_f$, must be the same in each case.

**b)** Ball (1) reaches the bottom first. If one considers only the vertical component of the motion, all of the balls start with the same velocity and move through the same height. Ball (1) has only one force acting on it, the gravitational force, and therefore has acceleration g. The other balls have a component of a another force upwards, in addition to the gravitational force downwards, so their net force vertically downwards is less than mg and their acceleration is less than g. Solving the kinematics equation $y_f - y_{=v_{iy}}t + ½ a_y t^2$, for example, for time, would demonstrate that ball (1) has the shortest time, because it has the highest acceleration and therefore Ball (1) arrives first. One can also argue that Ball (1) has the highest acceleration and the shortest distance to travel in the direction of motion, and therefore arrives first.

## C. Problem A Rubric

Analysis rubric for Problem A:

Part a) To be completely correct (C/C) on part (a), the answer must discuss conservation of energy in words or equations and include analysis of the initial and final conditions (in particular the fact that the distance traveled in the vertical direction is the same). If these two things are not explained clearly in words or equations, the answer is not counted as completely correct.

An example of a student's completely correct answer is:

All three balls reach the bottom with the same speed. In all three systems, energy is conserved, meaning the sum of potential and kinetic energy at one point equals the sum of potential and kinetic energy at another point. If we take the bottom to have a height = 0 and we can assume all three start at rest, then the initial potential energy, is the same for all three, since potential energy is mgh, is equal to the final kinetic energy. This means the final kinetic energy is the same for all three balls and the kinetic energy is $1/2\ mv^2$ and the velocities are all equal.

An answer was counted as partially correct (C/P), if the explanation is correct but lacks explicit detail, such as explicit analysis of the initial and final conditions or an explicit relationship between kinetic energy and velocity. The student's showed an understanding of the concepts and some instructors might have thought these answers to be sufficient, but they were counted as partial by our rubric.

Two examples of partially correct answers are:

(a) mgh = $1/2mv^2$, $v^2= 2gh$ All three will reach the bottom with the same velocity because of the above formula.
(b) They reach the bottom at the same speed. This is because they have the same $U_{grav}$ at the beginning with no KE and at the bottom it attains only KE and no $U_{grav}$. ME = KE + $U_{grav}$. So the $U_{grav}$ at the top = KE at the bottom.

The first is not completely correct, because the initial and final conditions, the fact that "h" is the same for all cases, is not explicitly discussed.

The second is not completely correct because the relationship between the velocity and the kinetic energy is not explicitly stated.

Examples of answers counted as correct choice, incorrect explanation (C/I) are below:

(a) They all have the same speed because they are falling due only to gravity and acceleration with the same mass.

(b) They are all the same because they start from the same height.

Examples of answers counted as Incorrect (I) are:

(a) In case 3, the ball reaches the bottom with the highest speed, because it has mg the momentum from the swing.

(b) Case 3 the ball has more distance to travel, so it has more time to build up speed.

Part b) To be counted as completely correct (C/C) in the part (b) of the problem, an answer must discuss the distance traveled and the acceleration of the ball in each of the three cases. A correct answer must include why the acceleration is higher or lower in each case. The answer could, for example, discuss the distance traveled in the y-direction being the same for each of the balls, but the acceleration being greater in case (1) because it has the greatest net force acting on it or analyze the distance and acceleration in the direction of motion in each case.

An example of a student's correct answer to version e:

If there was a way to have the balls fall at t = sqrt(2gh), then yes, but NO. The balls in 2 and 3 would take longer to fall than ball 1 because vertical forces from the normal force and tension would take away from the force of gravity. So if h is constant, than the force on the balls to travel the distance h would take longer.

Answers counted as partially correct (C/P) were answers that lacked detail. The statements were correct, but further information was needed, such as explicit discussion of the ranking of the accelerations based on the net forces acting or explicit reference to both acceleration and distance traveled.

An example of a student's partially correct answer to version f is:

No, case 1 will have the highest acceleration since there are no forces with a positive y-component acting on it. (Case 3 has the tension in the rope pulling up.) Since the acceleration of case 1 is greater, it will achieve its final velocity faster.

Examples of answers counted as correct choice, incorrect explanation (C/I) are:

(a) #1 It's the only direct path. Shortest distance.
(b) In case 1, the ball would reach the bottom first, it only has mass and gravity acting upon it while the others have things that delay them.

Examples of answers counted as Incorrect (I) are:

(a) 3 because it has the greatest momentum.

(b) In all three cases, the balls reach the bottom at the same time, because the force acting on all three is gravity.

**Appendix II**

**A. Problem B**

**Basic stem:**

For the figures below, all surfaces are frictionless. All masses are identical. All masses start from rest.

[See Figure 10]

**Version a (MC/FR):**

a) Choose the correct answer for the rank in order from greatest to least of the final velocities of the blocks. Explain why you chose the ranking you did. Show work and calculations.
  i)    D > B > A > C
  ii)   A = C > B = D
  ii)      D > A > B > C
  iii)     B > D > C > A
  iv)      B = D > A = C
  v)       A > D > C > B
  vi)      other ( show ranking and explain)

b) Choose the correct answer for the rank in order from greatest to least of the time it takes each block to reach the bottom. Explain why you chose the ranking you did. Show work and calculations.
  i)    D > B > A > C
  ii)   A = C > B = D
  iii)     D > A > B > C
  iv)      B > D > C > A
  v)       B = D > A = C
  vi)      A > D > C > B
  vii)     other ( show ranking and explain)

**Version b (FR):**

  a)  Rank in order from greatest to least the final velocities of the blocks. Explain your ranking. Show work and calculations.

  b)  Rank in order from greatest to least the time it takes each block to reach the bottom. Explain your ranking. Show work and calculations.

**Version c (FR Calculate):**

a) Calculate the final velocity of each of the blocks. Show your work.

b) Calculate the time it takes each block to reach the bottom. Show your work.

## B. Problem B Solution

A possible correct solution to Problem B (version b, Rank):

a) Using conservation of energy with the initial velocity and final height equal to zero, $v_i = 0$ and $h_f = 0$, we have $mgh_i = \frac{1}{2} mv_f^2$ and then $v_f = \sqrt{2gh}$. The final velocities can then be ranked by height and B = D > A = C is the correct ranking.

b) If we take our positive x-axis to be down the slope, the net force in that direction is $F = mg \sin\theta$, and by Newton's Second Law, $F = ma$, the acceleration down the slope is $a = g \sin\theta$. This can be combined with a kinematics equation, to solve for the time. For example, substituting $a = g \sin\theta$ into $x_f - x_i = v_i t + \frac{1}{2} at^2$, with $v_i = 0$, $x_f - x_i = d$, with d being the distance down the slope, one finds (using $\sin\theta = h/d$, where h is the initial height and not showing the algebraic process), $t = d\sqrt{2/gh}$. The time then depends both on the initial height, h and the distance down the slope, d. Substituting in the numerical values for h and d in each case, one finds A > D > C > B.

## C. Problem B Rubric

Part a) To be completely correct (C/C) on part (a), the answer must use either conservation of energy or a kinematics equation to arrive at $v_f = \sqrt{2gh}$, the relationship between the final velocity and the initial height, h. The student must then write out or choose the correct ranking.

As this is mostly an algebraic exercise, after the recognition of the use of conservation of energy or the correct use of a kinematics equation, we do not give examples of student's correct answers.

Answers counted as partially correct (C/P) predominantly consisted of solutions that had mathematical errors.

Answers counted as correct choice, incorrect explanation (C/I), were often arguments made based on angle, slope, distance down the slope, answers that were just statements or incomplete arguments.

An example of a C/I answer is:

B and D are equal, because B has half the distance of D but double the angle. This is the same for A and C.

Incorrect (I) answers consisted of arguments made based on angle, slope, distance down the slope, or answers that were just statements.

Examples of answers counted as Incorrect (I) are:

(a) B had the steepest incline, then D, then C, then A. The steepest slope would speed up the fastest, then D because its slope is the same as C, but it has a longer time to speed up. Then A, because it has the smallest slope and will travel slowly down it.

(b) A: 10m/s, B: 5m/s, C: 2 m/s, D: 5 m/s

Part b) To be counted as completely correct (C/C) in the part (b) of the problem, one needed to use two equations which could have been i) the equation that defines average velocity coupled with with $t = d/v_{avg}$, ii) a force equation and a kinematics equation, or iii) two kinematics equations.

There were no answers counted as partially correct.

The answers counted as correct choice, incorrect explanation (C/I) were almost exclusively students who wrote down the formula $t = d/v$, used the distance down the slope for d and the final velocity they had calculated in part (a) for v. This gives the correct ranking, but is not a correct way to work the problems.

Answers counted as Incorrect (I) consisted of arguments made based on angle, slope, distance down the slope, velocity or answers that were just statements.

Examples of answers counted as Incorrect (I) are:

(a) B, A = D, C. The ratio of the ramp length to height determines which will have the most time for the mass to reach the bottom.

(b) The slopes are a major factor in the time it takes. The greater the slope, the faster the time. A > C = D > B. (The student also showed calculation of each of the angles.)

**C. Problem Format Survey**

1) Please identify which type of physics exam or quiz question you prefer to be graded on:
   a) Multiple-choice
   b) true/false
   c) short answer
   d) show your work (including calculations) or explain your reasoning
   e) Other (describe)

Please discuss why you chose the answer you did.

2) Which type of physics exam or quiz question do you think is easier?
   a) Multiple-choice
   b) true/false
   c) short answer
   d) show your work (including calculations) or explain your reasoning
   e) Other (describe)

Please discuss why you think your answer is the easiest.

3) Which type of physics exam or quiz question do you think gives your instructor the most information on your understanding of the question being asked?
   a) Multiple-choice
   b) true/false
   c) short answer
   d) show your work (including calculations) or explain your reasoning
   e) Other (describe)

Please discuss why you chose the answer you chose.